\documentstyle[12pt,aaspp4]{article}

\begin{document}

\title{Evolution of Thick Accretion Disks Produced by Tidal Disruption Events}
\author{Andrew Ulmer\altaffilmark{1}}
\medskip
\affil{Princeton University Observatory, Peyton Hall, Princeton, NJ 08544}

\altaffiltext{1}{andrew@astro.princeton.edu}
 
\centerline{submitted to {\it ApJ}, August 1997}

\begin{abstract}

Geometrically thick disks may form after tidal disruption events,
and rapid accretion may lead to short flares followed by long-term,
lower-level emission.  Using a novel accretion disk code which relies
primarily on global conservation laws and the assumption that
viscosity is everywhere positive, a broad range of physically allowed
evolutionary sequences of thick disks is investigated.  The main
result is that accretion in the thick disk phase can consume only a
fraction of the initial disk material before the disk cools and
becomes thin.  This fraction is $\sim 0.5-0.9$ for disruptions around
$10^6$ to $10^7 M_\odot$ black holes and is sensitive to the mean
angular momentum of the disk.  The residual material will accrete in
some form of thin disk over a longer period of time.  The initial
thick disk phase may reduce the dimming timescale of the disk by a
factor of $\sim 2$ from estimates based on thin disks alone.
Assuming an $0.5~M_\odot$ initial thick disk, even if
the thin disks become advection dominated, the black hole mass to
light ratio can rise above $M_\odot/L_\odot = 1 $ in no less than
20 (0.1/$\alpha$) to 2000 (0.1/$\alpha$) years following a
tidal disruption event, depending on the mass of the black hole and
the initial conditions of the encounter.  The long-term emission will
be most prevalent around lower mass, $10^6 M_\odot$ black holes.
If the tidal disruption rates in these galactic nuclei
are $\sim 10^{-4} {\rm~yr}^{-1}$,
then about $10\%$ of the nuclei should exhibit the long-term
UV/optical emission at a level of $\sim 10^{38}
{\rm~ergs~s^{-1}}$.

\end{abstract}


\section{Introduction}

If black holes reside in the centers of galaxies, one
clinching sign of their existence would be the tidal disruption 
of stars when a star
passes within a tidal radius, $R_{\rm t} \sim 25 M_6^{-2/3} R_{\rm S}$,
of the black hole, where $M_6$ is the black hole mass in units of
$10^6 M_\odot$ and $R_{\rm S}$ is the Schwarzschild radius.
Tidal disruption events are
rare ($\sim 10^{-4}$ per year per $L_\star$ galaxy), bright
($L \sim L_{\rm Edd}$), short
(months to years), and seemingly unavoidable consequences
of $10^6-10^8 M_\odot$ black holes in galactic centers.
The general picture for tidal disruption of a star has been developed over
the last 20 years (e.g. Hills 1975; Young, Shields, \& Wheeler 1977;
Rees 1988).  The picture has evolved from one in which the disruption
rate was so high as to fuel quasars to one with a 
more modest rate in which disruptions are rare transients which may
nevertheless, slowly grow a lower mass, $\sim 10^7 M_\odot$, black hole
(Goodman \& Lee 1989).

At the time of disruption, half of the star is captured onto a range of
elliptical orbits around the black hole; the remainder is ejected from
the system.
Due to the relativistic effect of orbital precession,
debris orbits cross and may form strong shocks
(\cite{res88}).
The explicit evolution of the debris is complex and is dependent on the
spin and mass of the black hole as well as the properties of the star
and the stellar orbit.
As discussed by Kochanek (1994), at intersections between the debris streams,
energy and angular momentum may be transfered. The computational
complexities of the problem prohibit a direct numerical solution, but
as an approximation, a large fraction of the debris can be said to
circularize in a time,
\begin{equation}
\label{tcir}
t_{\rm cir} = n_{\rm orb} t_{\rm min},
\end{equation}
where $n_{\rm orb}$ is a number greater than 1, and probably between 2 and 10.
The minimum return time, $t_{\rm min}$ is the time for
the most bound material to return to pericenter, $R_{\rm p}$ is
\begin{equation}
\label{tmin}
t_{\rm min} = {2 \pi R_{\rm p}^{3} \over (GM_{\rm BH})^{1/2}
(2 R_\star)^{3/2} } \approx
0.11 \left( {R_{\rm p} \over R_{\rm t} } \right)^3
\left( {R_\star \over R_\odot} \right)^{3/2}
\left( {M_\star \over M_\odot} \right)^{-1}
M_6^{1/2}  {\rm years},
\end{equation}
where $R_\star$ and $M_\star$ are the radius and mass of the disrupted star.

The circularization process may produce a thick torus which would accrete on a
fast timescale.
Initially, the disk is expected to be geometrically
thick and radiation pressure supported, because of the large energy
difference ($\sim G M M_\star / 8  R_{\rm p}$)
between the nearly unbound elliptical orbits and a
subsequent circular orbit at $\sim 2 R_{\rm p}$, which is set by
conservation of angular momentum.
The disk is also highly optically thick with $\tau > 10^6$.
In comparison, our sun has $\tau \sim 10^{11}$ and a radius
$\sim 100$~times smaller
than the characteristic radius of a thick disk.
The thick disk's structure is very much like the envelopes of massive stars,
except that self gravity plays no role; the gravity in the disks is dominated
by the black hole.
The disk will remain geometrically thick if energy is dissipated
fast enough. In general, this energy dissipation rate
corresponds to an accretion rate which is super-Eddington.
The physical parameters of an tidal disruption encounter which produce
super-Eddington accretion are described in Ulmer (1997; hereafter U97).
Delineating the region in which thick accretion occurs is important not only
because super-Eddington tidal disruption events are the brightest
and therefore, the most readily observed flares, but also because one
expects the accretion of the stellar debris to be fastest when the disk
is very thick. In thin disk theory, the accretion time scales as
$\alpha^{-1} (z/r)^{-2}$, where $\alpha$ is the viscosity parameter,
$z$ is the height of the disk, and $r$ is the radial distance.
The thicker the disk, the faster it accretes.

The outline of the paper is as follows. In \S~\ref{secover}, we provide an
overview of the novel technique for forming evolutionary sequences out
of hydrostatic models of thick accretion disks.
The method relies primarily on global conservation laws
and the assumption that the viscosity is everywhere positive.
In \S~\ref{hydrostatic}, we describe the hydrostatic models which compose the
evolutionary sequence.
The method for evolving the models in time is presented in \S~\ref{evolsec}.
Initial conditions  are given in \S~\ref{initcond}
for the evolutionary sequences.
In \S~\ref{results}, we present results, including discussions of the
energy spectrum, the residual mass, and the importance of the
inner nozzle. Long-term emission properties of thin disks which form after
the thick disk phase are described in \S~\ref{longterm}. Conclusions
are given in \S~\ref{consec3}.

\section{Method Overview \label{secover}}

Here we investigate what ensues after the formation of
a thick disk
created from the debris of the tidal disruption of
a star by a massive black hole.
In such an event, it is believed that a fraction of the stellar debris will
be deposited in an accretion disk around the massive black hole.

The evolution of these thick, radiation-pressure-dominated
disks is studied as they generically expand,
accrete onto the black hole, and transfer angular
momentum outwards. Often, the thick disk expands, begins to
accrete, and after a fraction of the disk accretes onto the
black hole, the disk becomes thin. The evolution is followed until the
disks become thin.

The method is to construct hydrostatic models of thick
disks using a small number of parameters to specify the structure of the
disk. These models are then formed into a time sequence in which the
global quantities of mass, angular momentum, and energy are
conserved.
Viscosity is required to be positive, but not constant as a function
of radius or time.
The run of viscosity with radius is set by the time sequence of
models, as opposed to conventional methods in which the standard viscosity
parameter, $\alpha$, is
specified and taken as constant throughout the disk.
Time varying $\alpha$ has been commonly invoked in the study of dwarf-novae
(e.g. \cite{can88}, \cite{liv91})
as has radial variation in $\alpha$ (\cite{sma84}, \cite{can86});
this paper's method is well suited to this general case.
The range of allowable time sequences is investigated to determine
limits on relevant physical parameters such as the efficiency of
conversion of gravitational binding energy to radiation and
of the amount of disk mass left after the thick disk phase.
It is also possible to evaluate the importance
of initial values to the problem such as the
pericenter of the disrupted star and the mass of the black hole.
Our approach is similar to that of Abramowicz, Henderson, and Ghosh (1983)
who formed sequences of hydrostatic thick disk models, by
first assuming a constant viscosity parameter, $\alpha$, and then
evolving the disk according to global conservation laws.

The hydrostatic models for non-accreting disks
have four free parameters: the inner radius of the
disk, a polytropic constant which is related to the ratio of
gas pressure to radiation pressure, and two additional parameters which
specify the run of angular momentum with radius.
For accreting disks, it is necessary to specify only one parameter
for the run of angular momentum. The models are described in detail below.

\section{Hydrostatic Models \label{hydrostatic}}

For most the models we consider, the timescale for evolution of the disks
is much longer than the orbital time, so we construct hydrostatic models
(see, for example, Paczy\'nski and Wiita 1980, hereafter PW).
Neglecting the time dependence in the disk equations is a good approximation
unless the local viscous term, $\alpha$, approaches unity.
Additionally, there are non-hydrostatic effects of accretion at
the inner nozzle  which we discuss in \S~\ref{nozzle}.

To approximate the effect of the Schwarzschild geometry,
we use the gravitational potential described in (PW):
\begin{equation}
\label{pwpot}
\Psi = - {GM \over R - R_{\rm S} }, \hskip 1.0cm R_{\rm S} = {2GM \over c^2}.
\end{equation}
The Keplerian angular momentum in this potential is
\begin{equation}
\label{jkepeq}
j_{\rm Kep} = (GMR)^{1/2} { R \over R-R_{\rm S} }. 
\end{equation}
The Newtonian potential and angular momentum are
$-GM/R$ and $(GMR)^{1/2}$.
Both the Schwarzschild geometry and the PW potential differ from the
Newtonian potential
in that there is a marginally stable orbit, $R_{\rm ms} = 3 R_{\rm S}$, and
a marginally bound orbit, $R_{\rm mb} = 2 R_{\rm S}$.
The marginally stable orbit occurs because
angular momentum increases with decreasing radius when $ R < R_{\rm ms}$;
the binding energy, or the sum of the kinetic and potential energies,
reaches a minimum at $R_{\rm ms}$. The binding energy reaches zero at
$R_{\rm mb}$.
Keplerian thin disks can extend inwards only as far as $R_{\rm ms}$,
whereas thick disks can reach as far as $R_{\rm mb}$ because of
pressure gradients.
We make no further attempt to account for other general or special
relativistic effects, which we believe do not qualitatively 
affect any of the paper's results and may only slightly affect
the quantitative results.
(For example, the accretion efficiency of a thin disk is
5.7\% whereas the PW potential yields an efficiency of 6.3\%.)

The first free parameter of the hydrostatic
models is the inner radius, $r_{\rm i}$, where $r_{\rm i} > R_{\rm mb}$.
The structure of a thick disk is fully determined by the run of angular
momentum with radius, $j(r)$, at the surface.
In order for a disk to be dynamically stable angular momentum must increase
with radius, so we require the angular momentum distributions to
satisfy this constraint. 

For the models studied here, we take a modified power law angular
momentum distribution as a function with two parameters, $r_{\rm c} $ and
$b$:
\begin{eqnarray}
\label{angmom}
j(r)  = & A(r_{\rm c},b) r^{b-0.5} j_{\rm Kep}(r) &
{\rm for~}r > r_{\rm b} \\
j(r) = &  j(r_{\rm b})  & {\rm for~}r < r_{\rm b}, 
\end{eqnarray}
where the constant $r_{\rm b}/r_{\rm S} = 1/(1+b)$ ensures that
$ \partial j/\partial r \ge 0 $ for all $r$.
Examples of angular momentum distributions are shown in figure 1.
The slope, $b$, ranges between 0.5, where
$j(r)$ is Keplerian, and 0.0, where the angular momentum distribution is
flat.
The constant, $A$, is chosen so that the angular momentum is
equal to Keplerian at a specified central radius, $r_{\rm c}$.
For $r<r_{\rm c} $, the angular momentum is greater than Keplerian and
pressure gradients apply a force inwards.
For $r> r_{\rm c} $, the angular momentum is less than Keplerian and
pressure gradients apply a force outwards.

For hydrostatic thick disks which are (slowly) accreting, there is a natural
constraint on the angular momentum distribution
(Abramowicz, Jarosynski, and Sikora 1978).
The angular momentum at the inner radius, $r_{\rm i} $, must be Keplerian and
$r_{\rm i} \le 3 R_{\rm S} = R_{\rm mb}$.
For a non-accreting thick disk, the
inner radius can extend inwards of $R_{\rm mb}$ only as long as the
$j(r) > j_{\rm Kep}(r)$.
When $j(r)$ falls below $j_{\rm Kep}(r)$, orbits become unstable.

The last free parameter is related to the equation of state of the gas.
It is convenient to assume a polytropic equation of state.
Here we take
\begin{equation}
\label{poly1}
P = K \rho ^{4/3}, 
\end{equation}
where $K$, the polytropic constant, is taken to be the same everywhere in
the disk at a given instant of time. This approximation is explained
in~\S~\ref{addcon}.
The barytropic equation of state enforces uniform rotation on
cylinders.

Equations for hydrostatic equilibrium can be solved as a function
of the four parameters, $r_{\rm i,} r_{\rm c}, b,$ and $K$.
We adopt a convention of lowercase $r$ for cylindrical radial distance
and an uppercase $R$ for the spherical radial distance.

In the equatorial plane, the radial force balance equation is
\begin{equation}
\label{radforce}
{ 1 \over \rho } { \partial P \over \partial r } = 
 { \partial \Psi \over \partial r } + { v^2 \over r } , 
\end{equation}
where 
\begin{equation}
v = { j \over r }.
\end{equation}
The vertical force balance is
\begin{equation}
\label{vertforce}
-\sin (\theta) {\partial \Psi \over \partial R} + {1 \over \rho} {\partial
P \over \partial z} = 0. 
\end{equation}

The enthalpy, H(r), can be introduced and evaluated as
\begin{equation}
\label{entheq}
H(r) = \int { dP \over \rho } = 4 K \rho ^{1/3} = 4 { P \over \rho } , 
\end{equation}
for our equation of state (Eq. \ref{poly1}).
With equation \ref{radforce} this can be written,
\begin{equation}
\label{enthinteq}
H(r) =  \Psi (r) - \Psi (r_{i}) + \int _{r_{i}}^r { j^2 \over r^3 } dr.
\end{equation}

The differential equation \ref{vertforce} can be integrated to yield
\begin{equation}
\label{intvertforce}
\rho = \left[ {1 \over K^3} \right]
\left[{c_1\over 4} +{GM\over 4(R-R_{\rm S})}\right]^3, 
\end{equation}
where $c_1$ is constant on cylinders given by
\begin{equation}
\label{c1}
c_1 = 4K\rho_0^{1/3}(r)-{GM\over r-R_{\rm S}} = {H(r)} -
{GM \over r-R_{\rm S}},
\end{equation}
and $\rho_0$ is the density in the equatorial plane.

The disk surface is given by
\begin{equation}
\label{suface}
z_{\rm sur}(r) = \left\{ \left[{ r_{\rm i} -R_{\rm S} \over 1-
{(r_{\rm i} -R_{\rm S}) \over GM}
\int_{r_{\rm i}}^r j^2/r^3}
\right]^2 -r^2 \right\}^{1/2}
\end{equation}

With the density and run of angular momentum determined,
most physical quantities of the disk can be
obtained from integrations (either numerical or analytical).
The disk mass and angular momentum are
\begin{eqnarray}
M &=& \int_{r_{\rm i}}^{r_{\rm e}} {2 \pi r \Sigma } dr \\
J &=& \int_{r_{\rm i}}^{r_{\rm e}} {2 \pi r \Sigma j} dr 
\end{eqnarray}
The total disk energy is the sum of potential, kinetic, and internal energies (self-gravity is negligible):
\begin{eqnarray}
\label{eneqs}
E_{\rm kin}
&=& \int_{r_{\rm i}}^{r_{\rm e}} {2 \pi r \Sigma j^2 \over 2 r^2} dr \\
E_{\rm pot}
&=& \int_{r_{\rm i}}^{r_{\rm e}} 2 \int_{0}^{z_{max}} 2 \pi r {GM \rho \over R-R_{\rm S}}
dz dr \\
E_{\rm in}
&=& \int_{r_{\rm i}}^{r_{\rm e}} 2 \pi r (3-{3 \over 2} \beta)
2\int_0^{z_{max}} P(r,z) dz dr,
\end{eqnarray}
where $\beta$ is the ratio of gas pressure to total pressure and
is directly related to the polytropic constant, $K$.
The disk luminosity can be determined if we assume that the disk atmosphere,
which is radiation pressure dominated, 
is radiating critically like high-mass stars,
so they can be described by an Eddington model:
\begin{equation}
\label{eqrad1}
F_{\rm rad}= (1-\beta){c \over \kappa} g_{\rm eff},
\end{equation}
where $\kappa$ is the opacity per gram and $g_{\rm eff}$ is the
effective acceleration (including centrifugal force).
Integrating over the surface of the disk, the luminosity is (PW)
\begin{eqnarray}
\label{lumeqn}
L_{\rm rad} &=& 2\int_0^{2\pi} \int_{r_{\rm i}}^{r_{\rm e}} d\sigma \\
&=& (1-\beta){4\pi GM c\over \kappa} \int_{r_{\rm i}}^{r_{\rm e}}\left[
{j^4 R(R-1)^2 \over 
G^2M^2 z r^5 } - {2j^2 \over G M zr } + {rR \over z(R-R_{\rm S})^2} \right]dr,
\end{eqnarray}
where $\sigma$ is the element of surface area.
The luminosity may be altered by strong winds (e.g. U97)
as well as by effects in the funnel (e.g. \cite{nit82}), but for the
purposes of this paper we calculate the luminosity with Eq.~\ref{lumeqn}.
Figure 2 and figure 3 show many examples of these hydrostatic models.

\section{Evolution in Time \label{evolsec}}

Our goal is to form time sequences of hydrostatic models under four
physical constraints: global mass conservation,
global angular momentum conservation, 
global energy conservation, and positive local viscosity.
A computer code was written to perform the necessary numerical integrations
for mass, angular momentum, energy, etc.
Additionally, it imposes the physical constraints and forms time sequences
of models. The physical constraints greatly reduce the degrees of freedom in
the four dimensional parameter space as described below.

Mass and angular momentum conservation require that
for a non-accreting disk $M$ and $J$ are constant between models in
a time sequence. 
For an accreting disk, $M$ and $J$  must vary as
\begin{equation}
\Delta J = j(r_{\rm i}) \Delta M.
\end{equation}
As discussed above
$j(r_{\rm i}) = j_{\rm Kep}(r_{\rm i})$, for accreting disks.
The mass of the black hole, $M_{\rm BH} \gg M$, so
we take $M_{\rm BH} = {\rm const}$.
The code ensures that the constraints are meet to approximately one part in
$10^6$ per step which far exceeds the necessary accuracy as most
evolutionary sequences contain about 100 time steps.
The constraints of conserved mass and angular momentum are not necessarily met
if winds carry much the disk mass away (e.g. U97). If this is the
case, it will be very difficult to detect tidal disruption events from
their long-term emission, but the bright flares may be even more
optically accessible as the winds may reprocess some of the higher
energy photons to optical.

Energy is not conserved in the black hole/disk
system because of radiation.
However, we use energy conservation to determine the amount of
energy radiated between two models and thereby find the time step.
For a non-accreting disk,
\begin{equation}
\Delta t = {-\Delta E \over L},
\end{equation}
where $\Delta E$ is determined from equations \ref{eneqs} and
we impose the physical requirement that $\Delta E \le 0$.
For an accreting disk, the energy of the system changes
not only due to radiation but also due to mass loss on to the black hole.
In contrast to advection dominated models (e.g Narayan \& Yi 1995,
Chen~et~al. 1995)
the pressure at the inner radius falls to zero, so accreting matter has
negligible internal energy.
Therefore, the energy change between steps is
\begin{equation}
\Delta E = -E_{\rm rad} +\Delta M \left[ {-GM \over r_{\rm i} - R_{\rm S}} +
{j(r_{\rm i})^2 \over 2 r^2} \right].
\end{equation}

The constraint of positive viscosity ensures that angular momentum is
transfered outwards.
Specifically, for angular momentum distributions for which
$\partial\Omega / \partial r \le 0$ which are of general interest in
astrophysics, the
torque between two adjacent cylinders is always such that
angular momentum is flowing outwards on mass shells.
This constraint means that the torque between adjacent cylinders, $g$,
is greater than zero, so
$\dot{J}-\dot{M}j > 0$, since $\dot{J}=\dot{M}j +g$.
Thus, in contrast to the first three constraints, the
viscosity constraint needs to be applied locally.
In practice, the constraint is applied at twenty points in the disk
which are evenly spaced in mass and include the endpoints.

In geometric terms, the goal is to understand the physically allowable
paths that can be taken from an initial starting point in the 4-dimensional
parameter space of $r_{\rm i}$, the inner radius; $r_{\rm c}$,
the radius at which
the angular momentum is equal to the Keplerian value;
$b$, the slope of the angular momentum distribution;
and $K$, the polytropic constant.
These paths are discussed subsequently.

\subsection{Non-Accreting Evolution \label{nonacc}}

The physical constraints greatly limit the possible evolutionary
sequences. The two constraints of mass and angular momentum conservation in
the non-accreting case reduce the parameter space from four dimensions
to a 2-d surface (in 4-d) of models with constant mass and angular momentum.
The remaining two constraints from energy conservation and positive
viscosity further limit the possible evolutionary sequences, but in a
more complex manner.

For purposes of discussion, let us suppose that the energy loss from
radiation is negligible so that the disk evolves with constant energy.
Then, the allowable parameter space contracts from a surface to
a curve.
Finally, the requirement that viscosity is positive gives a direction
of evolution along the curve, so there is only one physically allowable
sequence of models which all lie along the curve.

Figure 4 shows evolution from a narrow torus of matter
to a flattened disk in this limit with no radiation loss.
Initially, when the angular momentum distribution is flat,
the disk's surface has a circular cross section for an infinitely narrow
torus as can be analytically demonstrated.
As the disk evolves, the circular cross section evolves into an
ellipse as the angular momentum distribution steepens.
The relationship between the slope of the angular momentum distribution,
$b$, and the ellipticity is derived analytically in
Appendix~B.
The numerical code well reproduces the relationship
predicted by the analytical approximation.
Like thin-disks, the non-accreting thick disk spreads
inwards and outwards in radius as time progresses.

In the more general and physically applicable case where energy is radiated,
the evolutionary sequences are not unique; however, they all lie on
the same 2-D surface.
The problem can be simplified because the disk is in general spreading both
inwards and outwards in radius until it begins to accrete, so the code may
can take steps at times when $r_{\rm i}$ is
equal to specified values less than the starting radius.
Once $r_{\rm i}$ is specified, there is only one degree of freedom left.
The range of possible models can be bracketed in a number of ways.
For instance, at each step one can take the minimum and maximum values of the
angular momentum slope, $b_{\rm min}$ and $b_{\rm max}$,
which provide solutions that meet the positive viscosity and energy
constraints.
Alternatively, the two limiting cases could be taken to be models,
denoted $\Delta E_{\rm min}$ and $\Delta E_{\rm max}$ in which
as little or as much energy is radiated as possible between timesteps.
Examples of these two limiting cases are shown in figure 5 and figure 6.
The two types of bracketing cases are identical for much
of the evolution of thick disks, which is to say
the limiting cases for $b_{\rm min}$ are the same as
for $\Delta E_{\rm min}$.
The case with $\Delta E_{\rm min}$ is one in which as little energy is
radiated as possible; and energy is eventually lost through low-efficiency
accretion. This limit is of physical interest, because it is the case which
determines the minimum accretion efficiency of the tidal disruption events
as well as determines the minimum amount of mass which remains
in the disk after the thick disk phase.
The $\Delta E_{\rm max}$ limit is one in which
the disk radiates as much energy as possible.
For the models we are investigating, the slope, $b$, increases
far enough that the disk eventually takes a Keplerian angular momentum
distribution and becomes essentially thin.
While thin disks are not well modeled by the approximations of the code
(particularly, the polytropic constant, $K$, is not be the same everywhere
in the disk), the late stages of evolution of a thin disk around a black hole
have been studied by Cannizzo, Lee, \& Goodman (1990; hereafter CLG),
and we can draw on their work to determine the subsequent evolution.

\subsection{Accreting Evolution}

While the disk is accreting,
there are only three free parameters as discussed above, because
the angular momentum must be Keplerian at the inner radius, $r_{\rm i}$.
This constraint creates an algebraic relationship between
$r_{\rm i}$, $b$, and $r_{\rm c}$.
Eliminating $r_{\rm c}$, for instance, reduces the parameter space to three
dimensions.

Because the disk is accreting, timesteps are taken at specified
values of $M$ less than $M_{\rm initial}$ in a similar fashion to how
timesteps were taken at given values of $r_{\rm i}$ in
the non-accreting case.
The mass and angular momentum constraints,
$\Delta M_{\rm n}=M_{\rm n-1}-M_{\rm n}$ and
$\Delta J_{\rm n} = j(r_{\rm i}) \Delta M_{\rm n}$,
can then be used to reduce the three dimensional space to a one-dimensional
space.
As for the non-accreting case,
we are left with the task of bracketing the possible evolutionary
tracks that are allowed within the constraints of energy loss and
positive viscosity.

One possible way to bracket the range of time sequences is to take extreme
values of $r_{\rm i}$ at each step.
This limit is computationally convenient because $r_{\rm i}$ is always between
2 and 3 $R_{\rm S}$ in the accreting case.
It is a physically interesting limit because the efficiency and
disk structure are strongly tied to the inner radius.
In particular, the total efficiency of the accretion process
determined by the effective potential at $r_{\rm i}$ and is given by
\begin{equation}
\label{epseq}
\epsilon ={1\over m_0 c^2}
\int_{ti}^{tf} L(t) dt \approx - {1\over m_0 c^2}
\left\{ \int^{t_{\rm f}}_{t_{\rm i}}
\left[- {GM_{\rm BH} \over r_{\rm i}(t) -
R_{\rm S} } + {j(t)^2 \over 2 r_{\rm i}(t)^2} \right]
{d M_{\rm d} \over d t} dt
\right\},
\end{equation}
where an initial disk of mass, $m_0$, accretes at a rate of
$\partial m/ \partial t(t)$ and radiates at a luminosity, $L(t)$
from time $t_{\rm i}$ to time $t_{\rm f}$.
As discussed in PW with reference to steady-state models,
because the effective potential (enclosed in brackets in Eq.~\ref{epseq})
rises from a minimum at $3R_{\rm S}$ to
zero at the marginally bound orbit, $2R_{\rm S}$, the more mass lost
at small $r_{\rm i}$, the lower the efficiency.
The full general relativistic treatment for a non-rotating black hole
has the same qualitative features and is quantitatively well
approximated by the potential of PW
(e.g. $\epsilon_{\rm max,S}/\epsilon_{\rm max,PW} \sim 1.1$).

\subsection{Additional Constraints \label{addcon}}

Additional constraints can be placed on the disk evolution
to make the evolution more physically realistic.
We discuss two constraints, one of which is related to the local viscosity,
and the second to the thermal timescale.

We may wish to place limits on the local or average value of the viscosity.
For instance when the strength of the viscosity is parameterized with
$\alpha$, i.e. $\eta = \alpha \eta_{\rm max} =
{2\over 3} \alpha  \rho v_{\rm s} z_0$, $\alpha$ should always be less
than 1. Additionally, when $\alpha$ approaches unity, neglect of the
time-dependent terms in the momentum-balance equations may become a poor
approximation.

Positive viscosity is enforced from timestep to timestep.
The viscosity varies with radius and time in contrast to most
approximations (e.g. standard $\alpha$-disk theory).
The local viscosity can be determined from the relation,
$\dot{J}=\dot{M}j +g$, since $\dot{J}$ and $\dot{M}$ (averaged over the
timestep) can be determined at each radius. The torque can be approximated
by $g \approx - \alpha 2\pi r^3 \Sigma v_{\rm s} 2 z_{\rm sur}
{\partial \Omega / \partial r}$.
In this manner, $\alpha$, can be determined as a function of time and radius.
At each timestep, additional constraints can be placed on $\alpha$.
For instance, $\alpha$ is required to be less than 1, and in some cases, the
mass weighted value of $\alpha$ is required to lie within a certain range.

A second constraint can be applied to enforce the
the assumption that the disk is a constant entropy polytrope.
The assumption requires that heat be exchanged rapidly within
the disk, because at large radii, the energy radiated is larger
than the energy locally dissipated (and vice-versa for small radii).
The energy radiated in a shell (Eqs.~\ref{eqrad1}-\ref{lumeqn})
goes as $r g_{\rm eff} \approx r^{-1}$.
The energy dissipated per unit volume is approximately
\begin{equation}
E_{\rm dis} \approx \left( r {d\Omega \over dr}\right)^2
\alpha \Sigma v_{\rm s}
\approx r^{2b-4} \alpha \Sigma v_{\rm s},
\end{equation}
which is generally a sharply falling function of radius, since $b$ is between
0 and 0.5, and $\Sigma$, $v_{\rm s}$, and even $\alpha$, generally
decrease with radius for $r > r_{\rm c}$.
Energy could be transported with the disk either through
circulation patterns which are limited by the sound speed, and which can be
very efficient as shown by contact binaries (see explanation by Tassoul
1992, and references therein) or
by radiative diffusion which is limited by the diffusion speed,
$\sim c/ \tau$, where $\tau$ is the optical depth between the points
of energy transfer. Radiative
diffusion may become the dominate heat transfer mechanism at very large
radii as the density falls.
Rather than model the radiative diffusion or a circulation
pattern, we consider two cases with the numerical code:
(1) the two mechanisms
together are efficient enough to keep the disk in a constant entropy state
or (2) the heat transfer is limited at a critical radius where the time
for heat to be transfered by circulation patterns exceeds the
Kelvin-Helmholtz time.
The transfer time is estimated as
\begin{equation}
t_{\rm trans}(r) \approx \int_{r_{\rm c}}^{r} {dr \over v_{\rm s}}.
\end{equation}
In the second case, we keep track of the amount of matter outside of
the critical radius and assume that it collapses to a thin disk on
the Kelvin-Helmholtz timescale. Once in a thin disk, the matter is
ignored for the remainder of the evolution.
In practice, there is very little difference between the main results
(\S~\ref{results}) in of both cases.


\section{Initial Conditions \label{initcond}}

We run the simulations for three black hole masses,
$10^6, 10^{6.5}, {\rm~and~} 10^7 M_\odot$ to cover the
range of black hole masses about which thick disks are likely to form.
Because  angular momentum is lost when the star disrupts,
the mean angular momentum of the initial disk is the same as that of
the star. If the debris circularizes into a narrow torus, then the
torus would have radius, $2 R_{\rm p}$, where $R_{\rm p}$ is the
pericenter of the star's orbit.
We consider disks with mean angular momentum corresponding to disruptions
at $0.5, 1, {~\rm and~ 2~} R_{\rm t}$, where $R_{\rm t}$ is the the tidal
radius.
The debris is taken initially to have constant specific angular momentum,
as might be expected since the debris does have constant angular momentum
when the star is disrupted and if
shocks efficiently circularize the debris, then energy may be more readily
transfered than angular momentum.
We begin the evolution with a high internal energy state, where
$r_{\rm i} \approx r_{\rm c}/2$, but the results are not very sensitive
to this condition ($\sim 10\%$ difference in the physical parameters discussed
below).
Nevertheless, there is some uncertainty
in the initial angular momentum state of the disk.

The simulations are most applicable to disruptions around black holes
with masses below $\sim 10^7 M_\odot$.
The simulations begin with a debris torus which subsequently accretes
at Eddington and super-Eddington accretion rates; however, the
marginally bound disruption debris only gradually returns and
circularizes on a timescale $t_{\rm min}$ (eq.~\ref{tmin}).
For black holes above $\sim 2 \times 10^7 M_{\odot}$, the circularization timescale
becomes so long that material enters the disk at a rate slower than required
to fuel the black hole at the Eddington rate (U97).
For these systems, thick disks cannot form.

For the lower mass black holes, even if the material is returned
at a high enough rate to fuel the black hole above the Eddington
limit, the accretion rate may be so high that much of the
material accretes before all of the debris is circularized (U97).
In this case, which appears to generally occur when $\alpha \gtrsim 0.01$,
much of the debris settles into the disk while the disk is accreting.
This returning matter cannot be directly accounted for with
the current simulations.

We address the problem with two types of runs. First, as described in
\S~\ref{addcon}, we can impose a restriction on the mass weighted
viscosity parameter
so that the accretion times are not so fast as to drain the disk before
the material circularizes. These are the runs labeled ``a'' in Tables~1-3.
Second, we can evolve a thick disk with as high viscosity as physically
possible,
even though in reality the disk could not have fully circularized. Such
sequences
may approximate the evolution when material is gradually fed to the disk,
because
such simulations have the same mean angular momentum, and it appears that
the angular momentum
is the most important parameter in the problem.
In these runs, the viscosity generally varies from near unity at
the innermost radius to a 0.1 at large radii.
These runs are marked ``c.''

\section{Results \label{results}}

The results from a variety of evolutionary tracks are shown in Tables~1-3.
The evolution is followed from the initial conditions given in
\S~\ref{initcond} to
a state in which the disk is no longer thick. An appropriate cutoff is when
the slope, $b$ becomes larger than 0.48, where 0.5 corresponds to a
Keplerian thin disk.
The evolutionary sequences are calculated with a variety of combinations
of the additional constraints of \S~\ref{addcon}
to investigate their effects on
the main results.

As explained in \S~\ref{nonacc}, 
in one type of physically allowed evolutionary
track, the disk cools quickly and forms a thin disk.
Thin disk evolution has been studied by CLG.
If the disk cools quickly ($\alpha$ is small),
there would likely be a relatively bright period of 1-3 years
when the disk shines below the Eddington luminosity, followed
by a self-similar phase in which the luminosity falls off as
$\sim t^{-1.2}$. The long-term evolution of disks is discussed
in \S~\ref{longterm}.

The main type of evolutionary track we study is one in which the
disk accretes quickly.
In these tracks, the maximum amount of mass is accreted onto the black hole
before the disks become thin.
The disks can accrete extremely fast, so if the viscosity is large enough
in the disks for such quick evolution, the accretion may be limited by
circularization of material which takes at least $t_{\rm min}$
(eq.~\ref{tmin}). Similarly, the efficiency of the accretion can become
extremely small in the idealized case where the all of the debris begins in
a circularized disk and it accretes at the maximum rate (``b'' and ``c'' in
tables 1-3).
The pericenter of the  stellar orbit is important in determining how much of
the disk can be accreted as can be seen by the strong correlation between
residual mass and pericenter, as is explained in \S~\ref{resmass}.
For the same reason, there is an inverse relationship between the black hole
mass and the residual mass.

\subsection{Spectra}

The spectra of the disks may be calculated at each time step
under the assumption that the surface is a thermal.
The optical depth is quite large, $\sim 10^6$, even to Thompson
scattering, so the radiation from the disk surface is most likely
thermal. With the the local luminosity specified by eqs~\ref{lumeqn},
the local temperature may be determined with the Stefan-Boltzmann law.
A Planck function is used to describe the local emission, and
is integrated over the disk surface to determine the final spectrum.
If as we believe, the scattering cross-section far exceeds
the absorption cross-section, the local spectrum could be
better described with a Wien peak and a modified black body spectrum
(e.g. Rybicki \& Lightman 1979). Given the ad~hoc nature of the angular
momentum distribution and uncertainties in the structure of
the inner funnel of the disk (e.g. \cite{nit82}),
such refinements are unwarranted.

Results of such calculations are presented in U97,
along with a description of current uncertainties which
may affect the energy spectrum.
As discussed in U97, the viewing angle of the disk is very important
because much of the emission is radiated from the inner regions of
the disk which are not visible to an edge-on viewer. The edge-on disk
is $\sim 50$ times less luminous than the face-on disk, although because
the outer disk is much cooler than the inner disk,
the difference seen in the optical bands may be significantly less, $\sim 5$.
Viewed face-on, the bolometric luminosities are a few times the
Eddington limit,
but the spectra are very hot, peaking between 50 and 100~eV.
The estimated 
bolometric corrections for U and V are $\sim 5.5$ and $7.5$, respectively.

\subsection{Residual Mass\label{resmass}}

In most cases 
a significant fraction of the mass of the thick disk
cools and forms a thin disk around the black hole, where it accretes
on a longer timescale (tables~1-3).
The main reason why mass cannot be fully accreted in the thick disk is
that the
specific angular momentum of the disk rises as the disk accretes,
so the disk must spread to large radii.
The specific angular momentum of material which accretes onto the black hole
is generally much smaller than the mean angular momentum of material
in the disk. For example, the disk formed by a solar-type
star with pericenter equal to the tidal radius which is disrupted by
a  $10^6 M_\odot$ black hole has mean specific angular momentum
$10 GM_{\rm BH}/c$,
whereas the angular momentum of the accreting material, has specific angular
momentum
\begin{equation}
{\sqrt{2} (r_{\rm i}/R_{\rm S})^{3/2}\over
(r_{\rm i}/R_{\rm S})-1 }{GM_{\rm BH} \over c} < 4 {GM_{\rm BH} \over c}.
\end{equation}
If $50\%$ of the disk accretes, then the average specific angular momentum of
the remaining material must increase to $\sim 16$, increasing the mean
distance of the disk as well as the accretion timescales for the remaining
material.
More generally,
if a fraction, $x$, of the disk accretes with a specific angular momentum,
$4 GM_{\rm BH} / c$, then
\begin{equation}
j_{\rm f} = {j_{\rm i} - x 4 GM_{\rm BH}/c  \over 1-x}.
\end{equation}
As the specific angular momentum of the disk grows, so does its
characteristic radius, $r \propto j_{\rm f}^2$ and the accretion timescale,
$t \propto r^{3/2} \propto j_{\rm f}^3$.
Therefore, when $90\%$ of the mass is accreted, the mass
accretion rate has fallen by $\sim 1000$.
Thick disks only exist in steady-state when the accretion rate is
super-Eddington, so
the disks will generally cool to form thin disks before accreting
a large fraction, e.g. $>90$\%, of the disk.
The exceptions in Table~3 occur when the initial angular momentum
(and therefore pericenter) of the star is small.
In fact, when the initial mean angular momentum of the disk approaches
that of the Keplerian angular momentum at marginally bound orbit,
the disk can gain very little specific angular momentum as it accretes.
The initial mean angular momentum will be this low when the encounters are
very close with $R_{\rm p} \approx 2.5$, which may not be uncommon
among disruption events around black holes with
$M \gtrsim 10^7 M_\odot$.

\subsection{Inner Nozzle \label{nozzle}}

While hydrostatic solutions have no inner nozzle, physically realistic
accreting solutions must have an opening at the inner radius.
Because of the importance of the inner radius in determining the efficiency of
the accretion, an understanding of the inner nozzle is vital.
In general, the structure of the inner nozzle depends on the
angular momentum at the inner radius, the mass accretion rate, and
the polytropic constant, K. 

The problem of an inner nozzle for thick disks
has been addressed in a number of ways.
For instance, Loska (1981), solves vertically integrated equations
under the assumptions of  vertical hydrostatic equilibrium,
that $z/r_{\rm i} \ll 1 $, and that the radial velocity of
matter is dependent only on the $r$-coordinate.
The critical point for the flow is extremely close ($< 10^{-4}r_{\rm i}$),
to the inner radius without the nozzle, for the solutions described by
Loska. The solutions have no simple scaling relation; however,
utilizing the fact that the critical point is extremely close to
the inner radius justifies a slight simplification of the problem which
provides scaling laws.

In appendix~B, we apply an approach similar to that of
Paczy\'nski and Sienkiewicz (1972) who addressed the Roche lobe
nozzle in contact binaries.
All of the physical quantities calculated in the code may be determined for
a disk with a nozzle,  with knowledge of the height of the nozzle
(half-thickness of the disk)  which obeys the scaling relation
(Appendix~B):
\begin{equation}
z_{\rm s} \propto \dot{M}^{1/8} K^{3/8}
r_{\rm i}^{5/16} (r_{\rm i}-1)^{7/8} M_{\rm BH}^{-1/4},
\end{equation}
so the dependences on physical parameters, especially mass accretion rates,
are weak.

The disks with nozzles differ only slightly from those without.
Even for the fastest accreting models, the residual mass (\S~\ref{resmass})
differs only by $\sim 10-20\%$. The main difference in the models
with inner nozzles is that the efficiency of accretion is reduced,
in some cases by a factor of two because kinetic energy is advected into the
black hole.

\section{Long-term Evolution \label{longterm}}

After the disk becomes thin, the evolution can no longer be tracked by
the thick disk evolutionary code.
To determine the long-term evolution,
we begin with the scaling relations derived by CLG for thin disks
which, in part, draw upon the self-similar
solutions of Pringle (1974).
The mass accretion rate onto the black hole
according to their scaling relations (their eqns 17 and 18) and
normalized to their numerical results is
\begin{equation}
\label{mdoteq}
\dot{m} \approx 1.5 \times 10^{25}
A^{7/8} \left({ \alpha\over 0.1 }\right)^{-1/4}
\left( { M_{\rm d}(0)\over 0.5 M_\odot}\right)^{7/8}
\left(M_{\rm BH} \over 10^7 M_\odot \right)^{5/24}
\left({ t\over 1 {\rm year}}\right)^{-19/16} {\rm ~gm~s}^{-1},
\end{equation}
where $M_{\rm d}(0)$ is the initial mass of the disk, and $A$ is the
ratio of mean angular momentum of the disk to mean angular momentum of a star
with pericenter equal to the tidal radius, and is $\sim 0.5-1.5$ initially.
If no angular momentum were carried into the black hole, $A M_{\rm d}(0)$ would
be constant as the disk accretes.
The results of \S~\ref{results} may be applied to the long term evolution by
using the final state of the thick accretion disk as the starting
point for the thin disk similarity solutions. The necessary parameters
are the mass and mean angular momentum of the residual disk.
Using Eq.~\ref{mdoteq}, both the luminosity as a function of time and
possible transitions to an less-efficient advection dominated
regime may be evaluated.
The evolution of the thin disk luminosity was determined by CLG. They
characterized the duration of the long-term {\em bolometric}
luminosity with the time, $t_1$, at which
$L_{\rm bol}/M_{\rm BH} = L_\odot/M_\odot$ and found it could 
be many thousand years.
Initially, the thin (and thick disks) are extremely hot and radiate only
a small fraction, $\lesssim 1\%$, of their luminosity in the optical
(see CLG and U97).
After 3000~years, CLG find that the peak of the broad
flux distribution shifts to 2000~\AA, so a much larger fraction of light
is emitted in the optical.
Assuming an efficiency of 6\% for conversion of mass to energy, as is
appropriate for thin disks around Schwarzschild black holes, the
disk will remain optically bright for (CLG)
\begin{equation}
\label{t1}
t_1  \approx 5000~f(t)^{16/19} A^{14/19} 
\left({ \alpha\over 0.1 }\right)^{-4/19} \left( { M_{\rm d}(0)\over
0.5 M_\odot}\right)^{14/19}
\left(M_{\rm BH} \over 10^7 M_\odot \right)^{-2/3}~{\rm years},
\end{equation}
where $f(t)$ which accounts for the relative brightness of stars and the
disk in the spectral band of choice; it is the ratio of the fraction
of disk light emitted in the band to the fraction of stellar light
emitted in the band, and is close to 1 in the optical after 3000 years.

We may now ask the question of how an early thick disk phase could affect
the long-term emission by applying Eq.~\ref{t1}
to the thin disks which remain after the thick disk phase.
The results are shown in Tables~1-3.
The timescales for dimming are extremely long.
If the tidal disruption rate in a typical galactic nuclei is
$\sim 10^{-4} {\rm~yr}^{-1}$ and the thin disks
accrete with efficiency $6\%$, then lower mass black holes would be
in a near constant state of emission.

If the accretion rate drops low enough, the accretion disk may become
advectively dominated. Advection would reduce the efficiency of the
accretion process below 6\% and could move much of the emission to a higher
temperature out of the optical band.
The disk may become advective when the mass accretion rate drops
below a critical value, $\dot{M}_{\rm crit} \approx
10~\alpha^2 L_{\rm Edd} / c^2$.
For a number of advective systems, $\dot{M}_{\rm crit} \approx
0.1 L_{\rm Edd} / c^2$ (e.g. Narayan 1996)
The time at which the accretion may become advective according to
Eq~\ref{mdoteq} is
\begin{equation}
\label{tadv}
t_{\rm adv} \approx 40~A^{14/19}  \left({ \alpha\over 0.1 }\right)^{-36/19} 
\left( { M_{\rm d}(0)\over 0.5 M_\odot}\right)^{14/19}
\left(M_{\rm BH} \over 10^7 M_\odot \right)^{-2/3}~{\rm years}.
\end{equation}
Below the critical mass accretion rate, the luminosity falls as
$\sim \dot{M}^2$. If the mass accretion rate continues to follow the
same form as Eq.~\ref{mdoteq}, then the point at which the bolometric
luminosity falls below $M_{\rm BH} L_\odot/M_\odot$ is
\begin{equation}
\label{t2}
t_2 \approx 400~A^{14/19}  \left({ \alpha\over 0.1 }\right)^{-20/19} 
\left( { M_{\rm d}(0)\over 0.5 M_\odot}\right)^{14/19}
\left(M_{\rm BH} \over 10^7 M_\odot \right)^{-2/3}~{\rm years}.
\end{equation}
Tables~1-3 shown that even in the advection dominated case, the
black holes will stay bright for an extended period of time.
For $10^7 M_\odot$ black holes, the disk will dim on the
timescale of $\sim 200 (0.1/\alpha)$ years, whereas
for $10^6 M_\odot$ black holes, the disk may remain bright 10 times as long
even when advection dominated. 

The self-similar solution does not account for the loosely bound
debris of the initial disruption which slowly rains down on the disk as
$\dot{M} \sim t^{5/3}$ (Rees 1988, Phinney 1989), but the return rate of the
marginally bound material falls below that of the accretion rate
from the self-similar solution within the first year.
We conclude that the long-term emission will be determined by the
evolution of the thin disk and not by return of marginally bound
material in agreement with CLG.

\section{Conclusions \label{consec3}}

Only a fraction of the disk mass can be consumed in the
initial thick disk phase, which is likely to occur for
tidal disruptions around $10^6-10^7~M_\odot$ black holes.
This fraction ranges from $\sim 0-0.99$ for disruptions around
$10^6$ to $10^7 M_\odot$ black holes, and has a typical value of
$0.5-0.9$.
The amount of the residual mass decreases for larger mass
black holes and increases with the pericenter of the disrupted star.
This residual material will accrete in a thin disk over a longer
period of time.
The existence of the initial thick disk phase may reduce the
dimming timescale of the disk by a factor of $\sim 2$ from estimates based
on thin disks alone.
Assuming that an $0.5 M_\odot$ initial thick disk,
even if the thin disks become advection dominated, the black hole mass to
light ratio can rise above $M_\odot/L_\odot = 1 $ in no less than
20 (0.1/$\alpha$) to 2000 (0.1/$\alpha$) years following a
tidal disruption event, depending on the black hole mass and
initial orbital of the disrupted star.
The long-term emission will
be most prevalent around $10^6 M_\odot$ black holes.
If the tidal disruption rates in these galactic nuclei
are $\sim 10^{-4} {\rm~yr}^{-1}$,
then at least $\sim 10\%$ of the nuclei should exhibit the long-term
UV/optical emission at a level of $\sim 10^{\rm 38} {\rm~ergs~s^{-1}}$.

\acknowledgements
I thank Jeremy Goodman, Bohdan Paczy\'nski, and David Spergel for
many helpful suggestions on this work.
AU was supported by an NSF graduate fellowship and NSF
grants AST93-13620 and AST95-30478.

\section{Appendix A \label{ellapp}}

In this appendix, the evolution of an infinitely narrow torus with
constant angular momentum is investigated.
This example can be evaluated
in large part analytically and therefore offers a check on the
the numerical analysis used elsewhere in the paper.

A thick disk with constant angular momentum and
barytropic equation of state of infinitesimal width in
a Newtonian potential has equipotentials (including the surface)
that are circular in cross-section. 
For example, at the surface of such a circular-cross-section disk,
\begin{equation}
z^2 +(r-(r_{\rm i}+\Delta))^2=\Delta^2,
\end{equation}
where the disk is centered on $(r_{\rm i}+\Delta)$ and the torus
has radius $\Delta = (r_{\rm e}-r_{\rm i})/2$.

More generally, we show that when the angular momentum is not
constant with radius, but rather a power law, the equipotentials
are ellipses.
At the surface of the disk (see section~\ref{hydrostatic})
\begin{equation}
z = \left\{ \left[{ r_{\rm i}-R_{\rm S} \over 1-{(r_{\rm i}-R_{\rm S})
\over GM} \int_{r_{\rm i}}^r j^2/r^3 dr}
\right]^2 -r^2 \right\}^{1/2}.
\end{equation}
For simplicity, the calculations are performed in a Newtonian potential,
($R_{\rm S} / r_{\rm i} \ll 1$),
the angular momentum is a power law with radius, ($j=A r^b$), and
the distance between the inner and outer radii is $2 \Delta$.

Within this framework, the equation for the surface is
\begin{eqnarray}
z & = & \left\{ \left[{ r_{\rm i}\over 1-{r_{\rm i} \over GM}
\int_{r_{\rm i}}^{r} A^2 r^{2b-3}dr}
\right]^2 -r^2 \right\}^{1/2} \\
z^2  &=& \left[{ r_{\rm i}\over 1-{r_{\rm i}^{2b-2} A^2 \over
GM (2b-3)} \left( ({r\over ri})^{2b-3} - 1 \right) } \right]^2 - r^2
\label{lasteq}
\end{eqnarray}

The amplitude of the angular momentum, $A$, is such that the angular momentum
is Keplerian at the center of the disk, $r_{\rm i}+\Delta$:
\begin{equation}
A = (r_{\rm i} +\Delta)^{1/2 -b} .
\end{equation}
The powers of A that are needed for the calculation, expanded to first order
in $\delta = \Delta / r_{\rm i}$ are
\begin{eqnarray}
A &=& r_{\rm i}^{1/2 -b} (1+ (1/2 -b) \delta) \\
A^2 &=& r_{\rm i}^{1 -2b} (1+ (1 -2b) \delta) \\
A^4 &=& r_{\rm i}^{2 -4b} (1+ (2 -4b) \delta) .
\end{eqnarray}
Writing $r$ as $r_{\rm i} + \epsilon$, where $-\Delta < \epsilon < \Delta$,
substituting into Eq.~\ref{lasteq}, and expanding to second order in
$\epsilon / r_{\rm i}$ and $\Delta / r_{\rm i}$ yields
\begin{equation}
z^2= r^2_{\rm i}
(1 + 2 {\epsilon \over r_{\rm i}} + 2(1-2 b) {\Delta \epsilon
\over r_{\rm i}^2} + 2 b {\epsilon^2 \over r_{\rm i}^2}) -r^2
\end{equation}
Replacing $\epsilon$ with $r -r_{\rm i}$ and expanding terms 
\begin{eqnarray}
z^2  &=& (-1 + 2b) ( r^2_{\rm i} + r^2 + 2 r \Delta
- 2 r r_{\rm i}  - 2 r_{\rm i} \Delta) \\
z^2  +(-1 + 2b) \Delta^2 &=& (-1 + 2b) ( r- r_{\rm i} - \Delta)^2 \\
{z^2 \over 1-2b} + ( r- r_{\rm i} - \Delta)^2 &=&  \Delta^2. \label{zbeq}
\end{eqnarray}
The shape of the surface (and equipotentials) ranges from
a circle when $b=0$ to a thin disk or line when $b=0.5$,
corresponding to Keplerian rotation.

As expected, the code utilized in earlier sections produces ellipsoidal
cross sections with the correct ellipticity.
As a non-accreting disk evolves, the inner-edge moves inwards
and the outer edge moves outwards. As the disk expands, it flattens.
Equation~\ref{zbeq} is a good description of disk equipotentials when
both $R_{\rm S}/r_{\rm i}$ and $(r_{\rm e}-r_{\rm i})/r_{\rm i}$ are small.

\section{Appendix B}

In this appendix, we derive a useful scaling relationship for
the thickness of the nozzle. 
We operate under the assumptions of Loska (1981), who
solved vertically integrated equations
under the following assumptions: the disk is
in vertical hydrostatic equilibrium,
$z/r_{\rm i} \ll 1 $,
and the disk is polytropic.
According to Loska, the critical point for the flow is extremely close
($< 10^{-4}r_{\rm i}$), to the inner radius without the nozzle, but
the solutions have no simple scaling relation.
The proximity of the critical point to
the inner radius justifies a slight simplification of the problem which
provides scaling laws.
We follow an approach similar to that of
Paczy\'nski and Sienkiewicz (1972) who addressed the Roche lobe
nozzle in contact binaries.

In absence of accretion, the disk forms a cusp at $r_{\rm i}$.
We expand the effective potential around this point, fixing $r=r_{\rm i}$:
\begin{equation}
\phi = \phi_0 [1 + a z^2] + {j^2 \over 2 r_{\rm i} } ,
\end{equation}
where the gravitational potential (PW) is
$\phi_0 = - {GM / (r_{\rm i}-R_{\rm S}) }$.
At the disk surface, the effective potential is
\begin{equation}
\phi_{\rm s} = \phi_0[1 + a z_{\rm s}^2] + {j^2 \over 2 r_{\rm i}^2},
\end{equation}
where $z_{\rm s}$ is the half-thickness of the nozzle, and
\begin{equation}
a= { 1 \over r_{\rm i} (r_{\rm i} - R_{\rm S}) }.
\end{equation}

In hydrostatic equilibrium,
\begin{equation}
\label{hydroeq}
{dP \over \rho} = - d\phi.
\end{equation}
Integrating eq.~\ref{hydroeq} at $r=r_{\rm i}$, with
a polytropic equation of state, $P = K \rho^{1 + 1/n}$, yields
\begin{equation}
\label{h2}
(n+1) K \rho^{1/n} + \phi = \phi_{\rm s, hyd},
\end{equation}
where $(n+1) K \rho^{1/n} = H(r)$ is the enthalpy (eq.~\ref{entheq}),
and $\phi_{\rm s, hyd}$ is the effective potential evaluated at the
the hydrostatic disk's surface.
When the material has a constant radial velocity, but maintains its
barytropic equation of state, eq.~\ref{h2} becomes a Bernoulli equation:
\begin{equation}
\label{berno}
{1\over 2} v^2 + (n+1)K \rho^{1/n} + \phi = {\rm const}
\end{equation}
If the location of the surface were known, the enthalpy in the equatorial
plane can be determined at
$r_{\rm i}$ (since the value of $\rho$ in the
equatorial plane is known from vertical force balance
(eqs.~\ref{vertforce},\ref{intvertforce},\ref{c1})).
Once determined at $r_{\rm i}$, the enthalpy can be found throughout the plane
(eq.~\ref{enthinteq}) from which all other physical quantities can be
derived.
The value of the enthalpy at $r_{\rm i}$ is
\begin{equation}
H(r_{\rm i}) = {GM_{\rm BH} \over r_{\rm i} - R_{\rm S}} -
{1 \over (r_{\rm i}^2 + z_{\rm s}^2)^{1/2} - R_{\rm S}}.
\end{equation}

The thickness of the nozzle may be estimated as follows.
The flux of matter through the nozzle is limited by 
\begin{equation}
F = 4 \pi r_{\rm i} \int_0^{z_{\rm s}} \rho v_{\rm x} dz \lesssim
4 \pi r_{\rm i} \int_0^{z_{\rm s}} \rho v dz.
\end{equation}
The maximum value of $\rho v$ may be determined by differentiating
eq.~\ref{berno}:
\begin{equation}
v dv + K{n+1 \over n} \rho^{1/n -1} d\rho = 0,
\end{equation}
and using the constraint that at maximum
\begin{equation}
d(\rho v ) =0 = \rho dv + v d\rho.
\end{equation}
Combining the last two equations with the Bernoulli equation yields
\begin{equation}
(\rho v)_{\rm max} = \left( {\phi_{\rm s} - \phi \over n+ 0.5} \right)^{n+0.5}
\left( {n \over K( n+ 1)} \right)^n.
\end{equation}
The flux is then limited to
\begin{eqnarray}
F_{\rm max} &=& 4 \pi r_{\rm i} \int_0^{z_{\rm s}}
\left( {\phi_{\rm s} - \phi \over n+ 0.5} \right)^{n+0.5}
\left( {n \over K( n+ 1)} \right)^n dz \\
 &=& 4 \pi r_{\rm i} 
\left( {- \phi_0 a \over n+ 0.5} \right)^{n+0.5}
\left( {n \over K( n+ 1)} \right)^n \int_0^{z_{\rm s}}
(z_{\rm s}^2 - z^2)^{n+1/2} dz \\
&=& 4 \pi r_{\rm i} 
\left( {- \phi_0 a \over n+ 0.5} \right)^{n+0.5}
\left( {n \over K( n+ 1)} \right)^n z_{\rm s}^{2n+2} C_1,
\end{eqnarray}
where
\begin{equation}
C_1= {\sqrt{\pi} \Gamma(1.5+n) \over 2 \Gamma(2+n)}
= .429515 {\rm ~for~n=3}
\end{equation}
The true matter flux, $\dot{m}$ is slightly less than $F_{\rm max}$,
but in the case of Roche overflow in binaries, the difference is no
more than 20\%, a
negligible amount considering the crudeness of the assumptions.
Nevertheless, a direct
calibration on $\dot{m}/F_{\rm max}$ with numerical simulations would
be welcomed.
Simply equating $\dot{m}$ and $F_{\rm max}$ gives an estimate of the
nozzle half-thickness
\begin{equation}
z_{\rm s} = \left[ {\dot{M} K^n \over r_{\rm i}
(-\phi_0 a)^{n+1/2} } \right]^{1/(2n+2)}
\left[ {1\over 4 \pi} {n+1 \over n}^n (n+1/2)^{n+1/2} {1 \over C_1}
\right]^{1/(2n+2)}
\end{equation}

For $n = 3$, 
\begin{eqnarray}
z_{\rm s}& =& 1.56084
\left[ {\dot{M} K^3 \over r_{\rm i}
(-\phi_0 a)^{7/2} } \right]^{1/8} \\
&\propto & \dot{M}^{1/8} K^{3/8}
r_{\rm i}^{5/16} (r_{\rm i}-1)^{7/8} M_{\rm BH}^{-1/4}
\end{eqnarray}

\newpage




\newpage

\begin{table}
\caption{  $M_{\rm BH} = 10^6$.
Thick disk simulations with (1) the pericenter, $R_{\rm p}$, in units of the
tidal radius, $R_{\rm t}$,
(2) the total time the thick disk was evolved, in seconds, (3) residual mass
after thick disk phase as a fraction of the initial mass of
$0.5~M_\odot$,
(4) the efficiency of accretion (eq.~\ref{epseq}), (5) the time, $t_1$, for
a thin disk with 6\% accretion
efficiency to reach $L_{\rm bol}/M_{\rm BH} = L_\odot/M_\odot$ assuming
self-similar evolution discussed
in \S~\ref{longterm}, (6) the time, $t_{\rm adv}$, at which the thin disk
could shift to an advection
dominated mode if $\alpha \sim 0.1$, (7) the time, $t_2$, for a thin disk
which is advection dominated with
$\alpha \sim 0.1$ to reach $L_{\rm bol}/M_{\rm BH} = L_\odot/M_\odot$.
Runs labeled ``a'' have a restricted viscosity parameter, $\alpha$
(generally $\lesssim 10^2$), such that the
disk does not accrete faster than the debris can circularize.
Runs labeled ``b'' allow for the outer disk to cool and become thin as
described in \S~\ref{addcon} and allow the maximum viscosity.
Runs labeled ``c'' allow the the maximum viscosity ($\alpha=1$
and therefore the
least efficient accretion.
In runs labeled ``d,'' the disks cooled and became thin before any
accretion took place.
\label{tab1}}
\begin{tabular}{lcllllll}
$R_{\rm p} (R_{\rm t})$ &
$t_{\rm tot}$  &
$m_{\rm res}$ 
& effic &
$t_{1}$ & $t_{\rm adv}$ & $t_2$ &
notes  \\
\hline
 2  &                 &     &         & $3.0\times 10^{4 }$& $2.4\times 10^{2 }$& $2.4\times 10^{3 }$&a,d \\
 2  & $3.5\times10^6$ & 0.42& 0.012   & $2.8\times 10^{4 }$& $2.2\times 10^{2 }$& $2.2\times 10^{3 }$&b \\
 2  & $9.6\times10^6$ & 0.18& 0.019   & $2.6\times 10^{4 }$& $2.0\times 10^{2 }$& $2.0\times 10^{3 }$&c \\
 1  & $1.6\times10^7$ & 0.52& 0.033   & $2.3\times 10^{4 }$& $1.8\times 10^{2 }$& $1.8\times 10^{3 }$&a  \\
 1  & $2.9\times10^6$ & 0.34& 0.0083  & $1.9\times 10^{4 }$& $1.6\times 10^{2 }$& $1.6\times 10^{3 }$&b \\
 1  & $7.1\times10^6$ & 0.12& 0.014   & $1.8\times 10^{4 }$& $1.4\times 10^{2 }$& $1.4\times 10^{3 }$&c \\
 0.5& $9.7\times10^6$ & 0.26& 0.016   & $1.5\times 10^{4 }$& $1.2\times 10^{2 }$& $1.2\times 10^{3 }$&a  \\
 0.5& $3.0\times10^6$ & 0.22& 0.0063  & $1.3\times 10^{4 }$& $1.0\times 10^{2 }$& $1.0\times 10^{3 }$&b \\
 0.5& $6.0\times10^6$ & 0.08& 0.01    & $1.2\times 10^{4 }$& $9.4\times 10^{1 }$& $9.4\times 10^{2 }$&c \\
\end{tabular}
\end{table}

\begin{table}
\caption{ $M_{\rm BH} = 10^{6.5}$. For description, see table~1.
\label{tab2}}
\begin{tabular}{lcllllll}
$R_{\rm p} (R_{\rm t})$ &
$t_{\rm tot}$  &
$m_{\rm res}$ 
& effic &
$t_{1}$ & $t_{\rm adv}$ & $t_2$ &
notes  \\
\hline
 2  &                 &     &         & $1.4\times 10^{4 }$& $1.1\times 10^{2 }$& $1.1\times 10^{3 }$&a,d \\
 2  & $2.0\times10^6$ & 0.35& 0.015   & $1.2\times 10^{4 }$& $9.8\times 10^{1 }$& $9.8\times 10^{2 }$&b \\
 2  & $1.6\times10^5$ & 0.20& 0.015   & $1.1\times 10^{4 }$& $8.5\times 10^{1 }$& $8.5\times 10^{2 }$&c \\
 1  & $8.5\times10^6$ & 0.56& 0.037   & $9.1\times 10^{3 }$& $7.3\times 10^{1 }$& $7.3\times 10^{2 }$&a  \\
 1  & $1.5\times10^6$ & 0.25& 0.0093  & $7.8\times 10^{3 }$& $6.3\times 10^{1 }$& $6.3\times 10^{2 }$&b \\    
 1  & $2.6\times10^6$ & 0.13& 0.013   & $7.0\times 10^{3 }$& $5.6\times 10^{1 }$& $5.6\times 10^{2 }$&c \\
 0.5& $3.3\times10^6$ & 0.21& 0.011   & $4.6\times 10^{3 }$& $3.7\times 10^{1 }$& $3.7\times 10^{2 }$&a  \\   
 0.5& $1.0\times10^6$ & 0.11& 0.0048  & $4.1\times 10^{3 }$& $3.3\times 10^{1 }$& $3.3\times 10^{2 }$&b \\    
 0.5& $1.4\times10^6$ & 0.08& 0.0068  & $4.3\times 10^{3 }$& $3.5\times 10^{1 }$& $3.5\times 10^{2 }$&c \\     
\end{tabular}
\end{table}

\begin{table}
\caption{ $M_{\rm BH} = 10^7$. For description, see table~1.
\label{tab3}}
\begin{tabular}{lcllllll}
$R_{\rm p} (R_{\rm t})$ &
$t_{\rm tot}$  &
$m_{\rm res}$ 
& effic &
$t_{1}$ & $t_{\rm adv}$ & $t_2$ &
notes  \\
\hline
 2  &                 &     &         & $6.7\times10^4$ & $5.4\times 10^{1 }$& $5.3\times 10^{2 }$&a,d \\    
 2  & $1.0\times10^6$ & 0.25& 0.015   & $4.6\times10^4$ & $3.7\times 10^{1 }$& $3.7\times 10^{2 }$&b \\     
 2  & $1.1\times10^6$ & 0.21& 0.016   & $4.5\times10^4$ & $3.6\times 10^{1 }$& $3.6\times 10^{2 }$&c \\    
 1  & $5.3\times10^6$ & 0.64& 0.056   & $4.3\times10^4$ & $3.5\times 10^{1 }$& $3.5\times 10^{2 }$&  \\
 1  & $5.5\times10^5$ & 0.11& 0.0060  & $2.3\times10^4$ & $1.9\times 10^{1 }$& $1.9\times 10^{2 }$&b \\
 1  & $5.6\times10^5$ & 0.10& 0.0061  & $2.3\times10^4$ & $1.8\times 10^{1 }$& $1.8\times 10^{2 }$&c \\  
 0.5& $5.5\times10^5$ & 0.05& 0.0024  & $5.8\times10^3$ & $4.7\times 10^{0 }$& $4.7\times 10^{1 }$&  \\  
 0.5& $2.9\times10^4$ & 0.01& 0.00045 & $3.0\times10^3$ & $2.4\times 10^{0 }$& $2.4\times 10^{1 }$&b \\  
 0.5& $5.9\times10^5$ & 0.01& 0.00038 & $2.6\times10^3$ & $2.1\times 10^{0 }$& $2.1\times 10^{1 }$&c \\  
\end{tabular}
\end{table}

\begin{figure}
\plotone{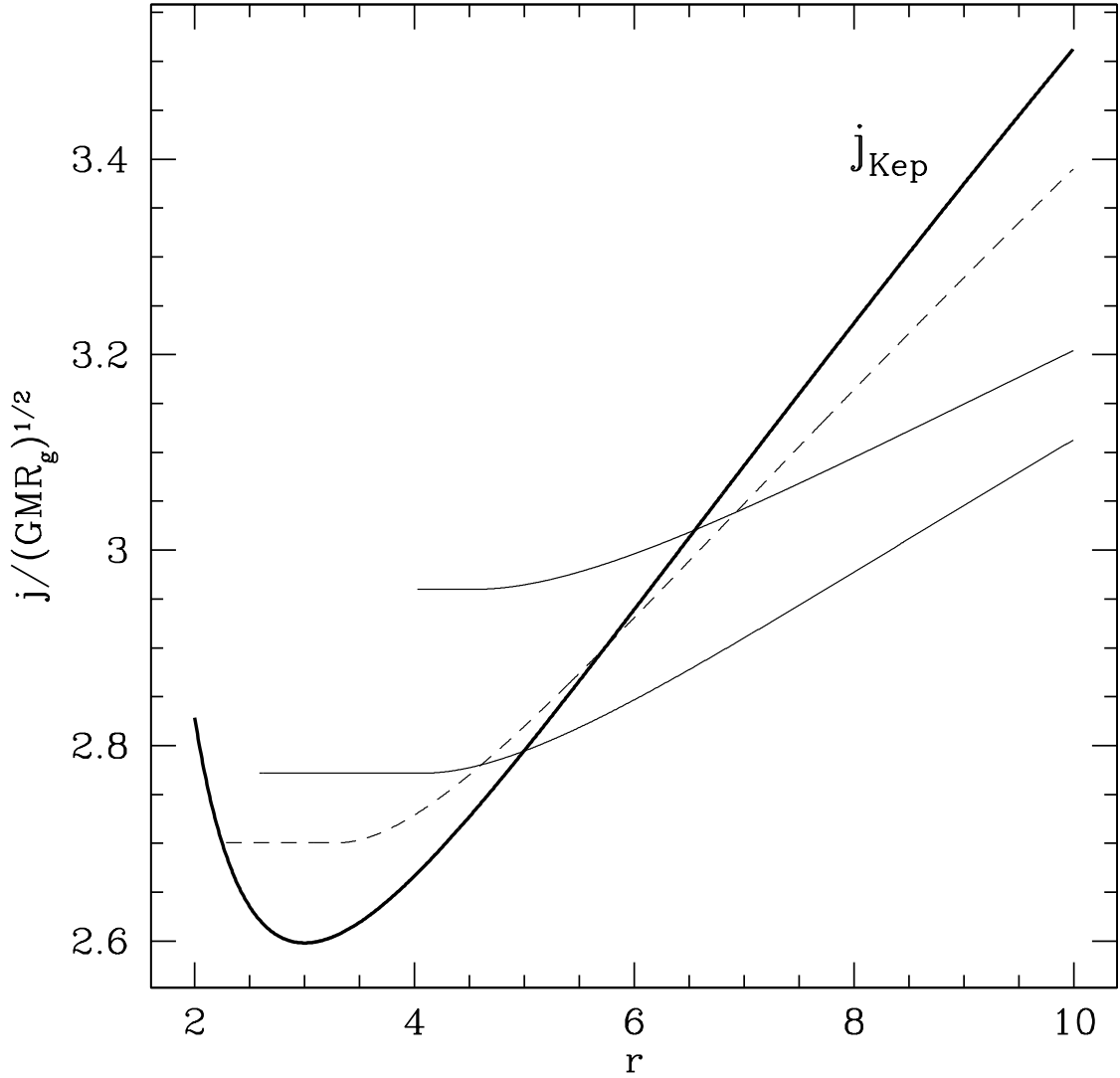}
 \caption{Angular momentum as a function of radius for a Keplerian disk
(thick solid line; eq.~\ref{jkepeq}) in the pseudo-Newtonian
potential of PW, (eq.~\ref{pwpot})
for non-accreting disks (thin solid lines), and for an accreting disk
(dashed line).}
\end{figure}

\begin{figure}
\plotone{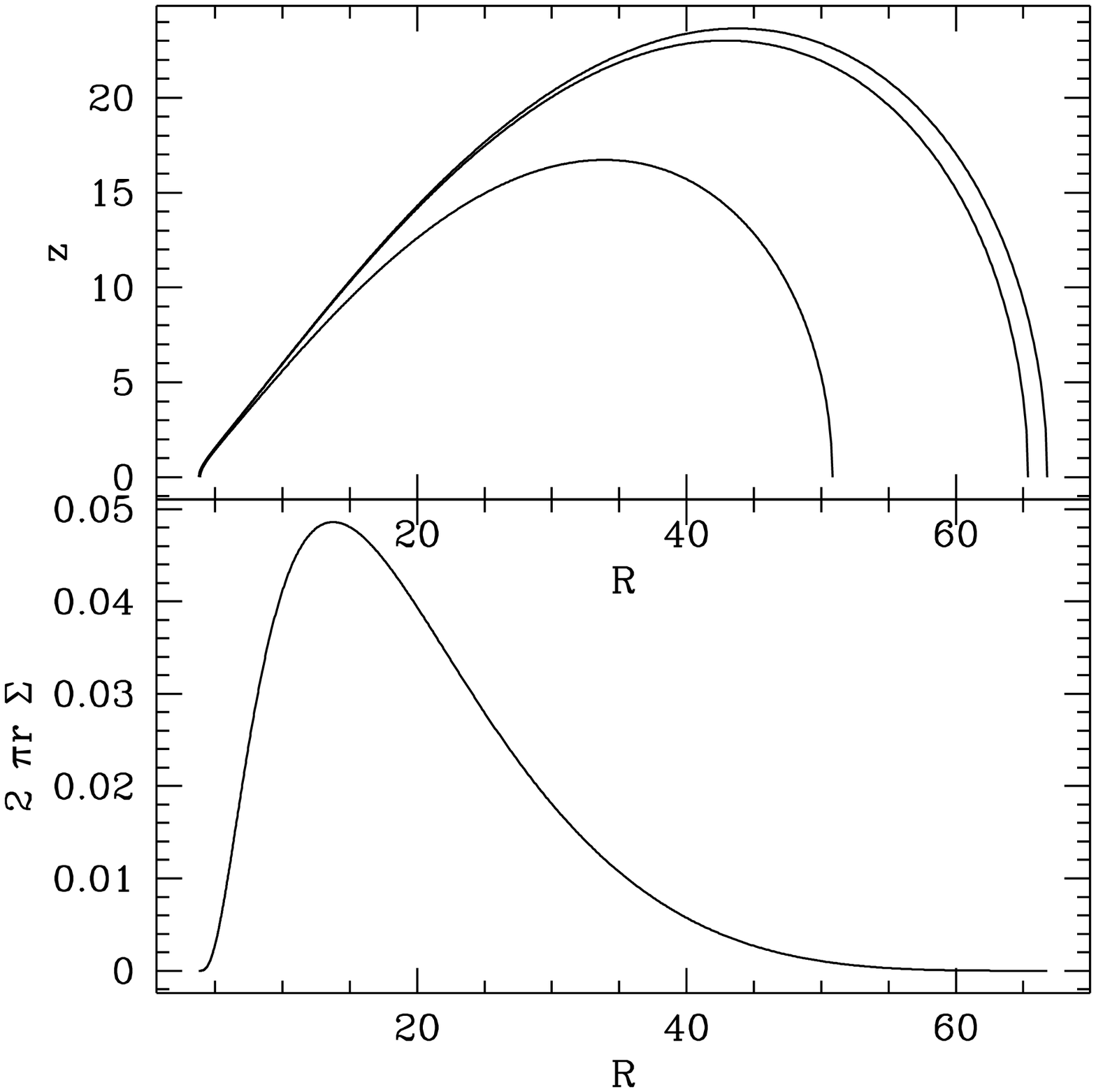}
 \caption{A non-accreting thick disk.
The top panel shows equipotentials containing 100 \% (surface),
95 \%, and 50 \% of the mass in the r/z plane.
The bottom panel shows the mass distribution as a function of radius.}
\end{figure}

\begin{figure}
\plotone{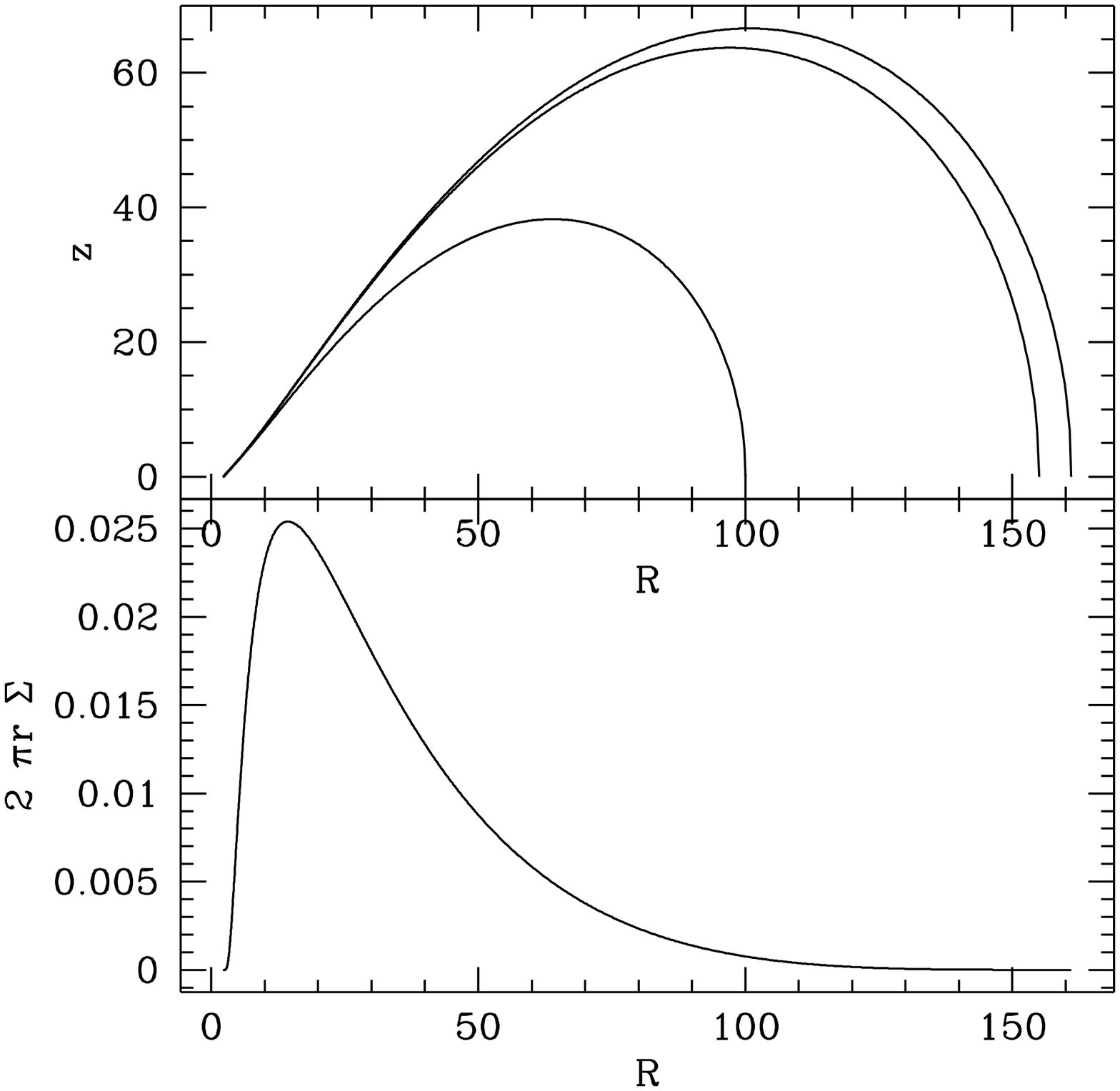}
 \caption{An accreting thick disk.
The top panel shows equipotentials containing 100 \% (surface),
95 \%, and 50 \% of the mass in the r/z plane.
The bottom panel shows the mass distribution as a function of radius.}
\end{figure}

\begin{figure}
\plotone{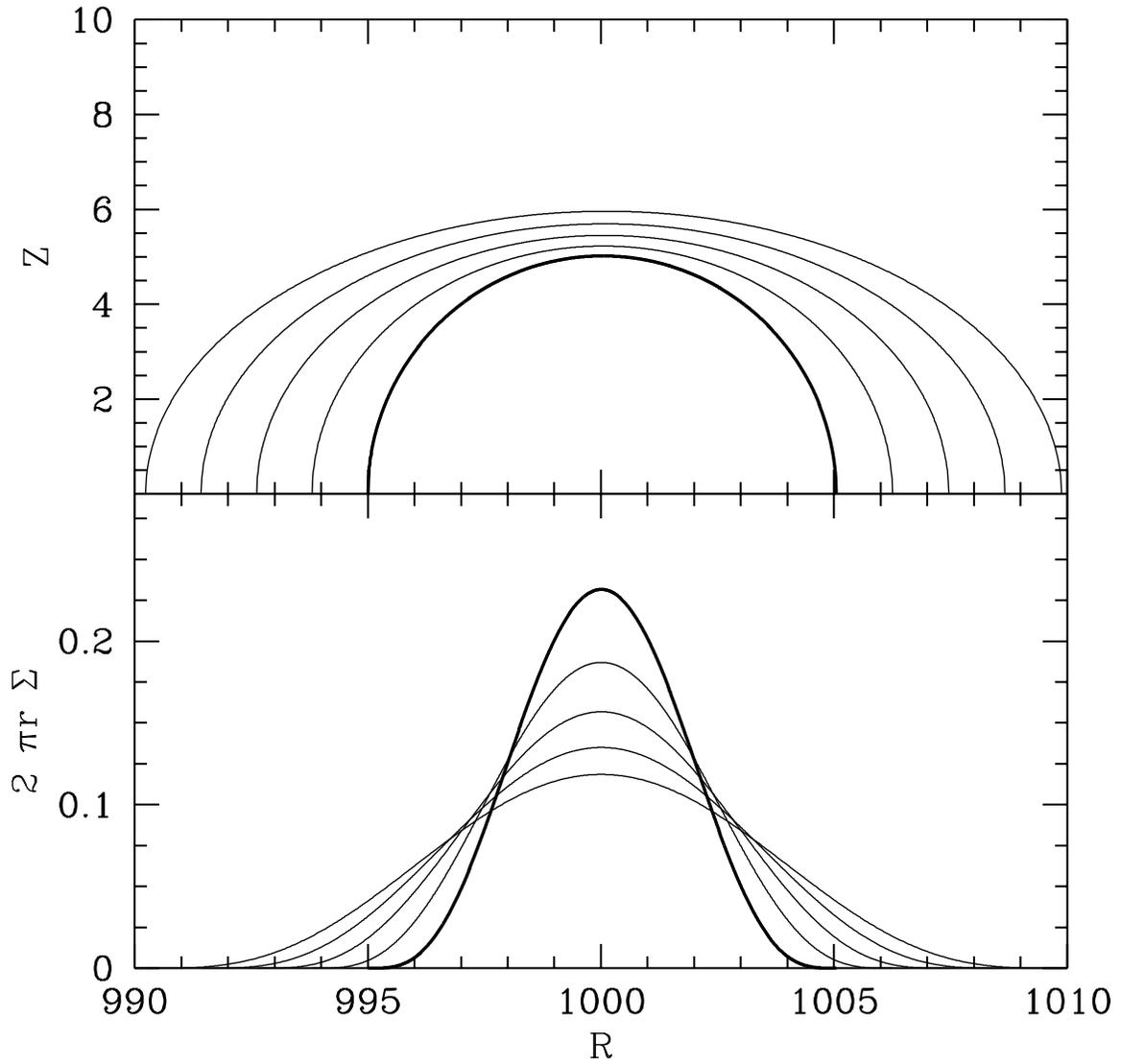}
 \caption{Evolution of a circular tube into a disk using the
evolution code. 
The code well reproduces the analytical results
of Appendix~B.
The top panel shows cross-sections of the surface at different times.
The bottom panel shows the mass distribution as a function of radius.}
\end{figure}

\begin{figure}
\plotone{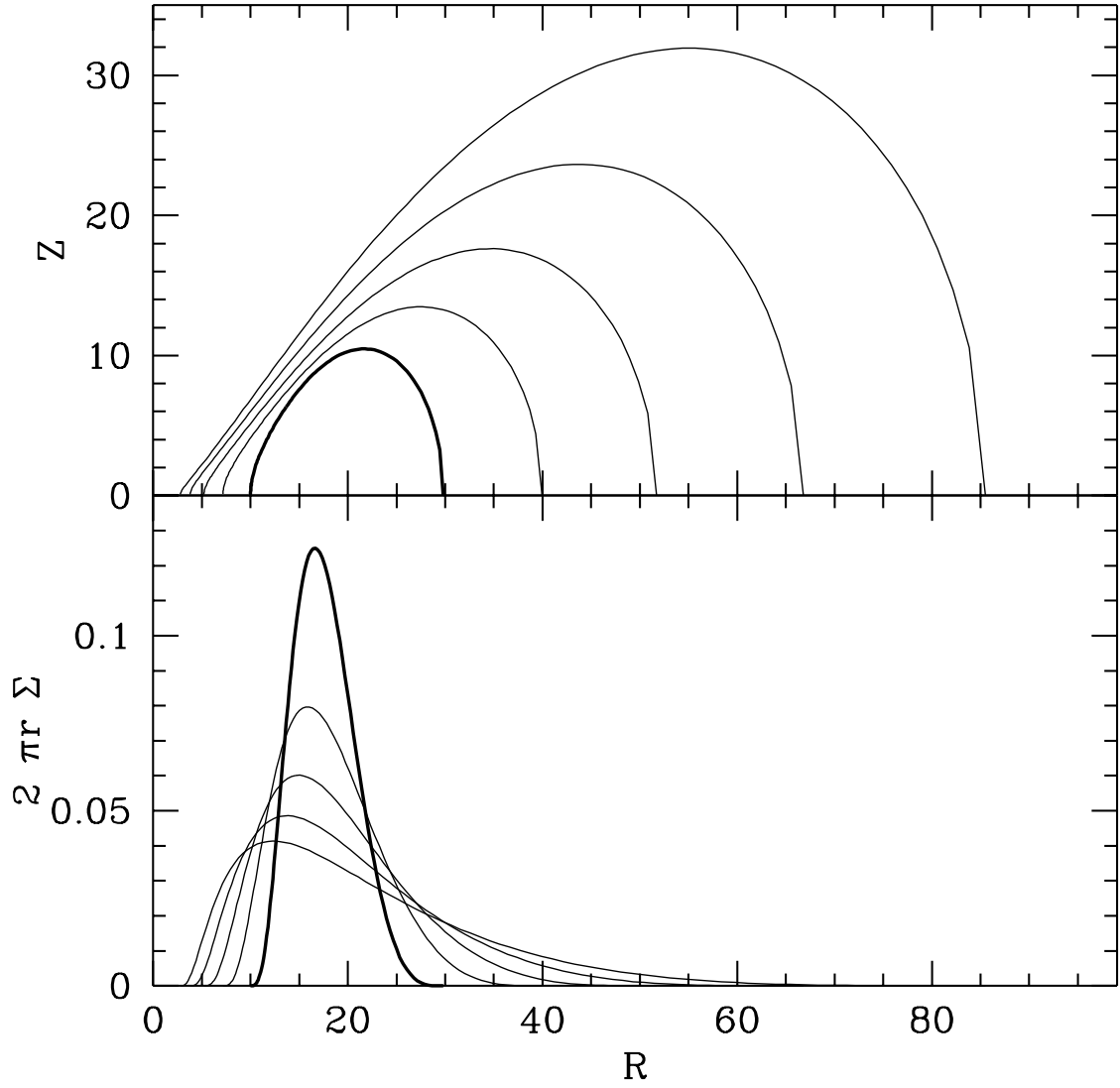}
 \caption{Evolution of a non-accreting disk in the limit that very little
energy is radiated.
The initial model is shown as a thicker line.
The different curves correspond to the disk at different times as it expands.
}
\end{figure}

\begin{figure}
\plotone{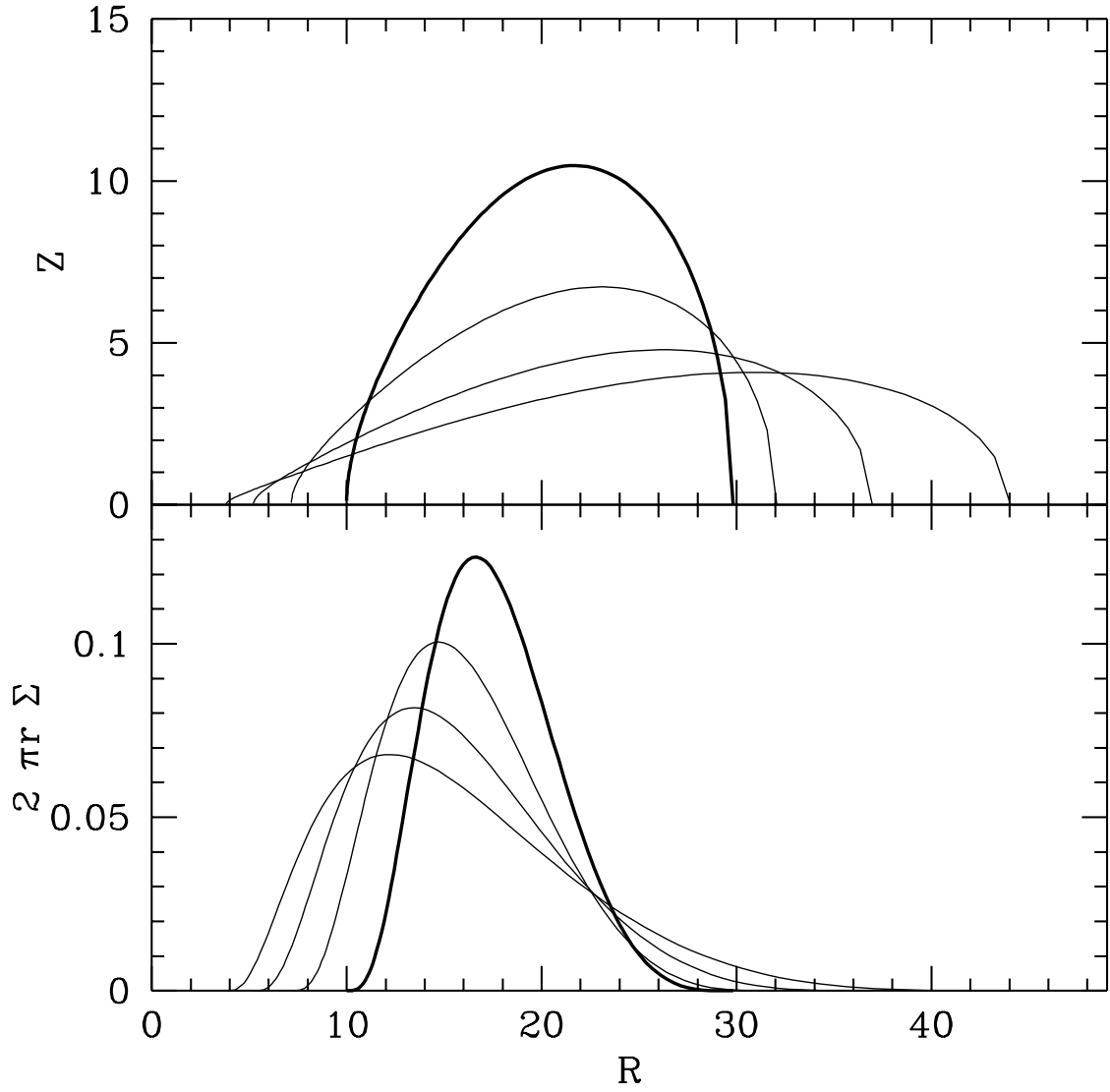}
 \caption{Evolution of a non-accreting disk in the limit that the maximum
energy is radiated at each step, so that the disk quickly becomes thin.
The initial model is shown as a thicker line.
The different curves correspond to the disk at different times as it expands.
}
\end{figure}

\end{document}